\documentclass[aps,pra,showpacs,twocolumn]{revtex4}
\usepackage{graphicx}
\usepackage{dcolumn}
\usepackage{bm}
\usepackage{color}
\usepackage{amsmath}
\usepackage{amssymb}
\usepackage{setspace}

\begin{document}
\title{Monotonic quantum-to-classical transition enabled by positively-correlated biphotons}
\author{Rui-Bo Jin$^{1}$}
\email{jrbqyj@gmail.com}
\author{Guo-Qun Chen$^{1}$}
\author{Hui Jing$^{2}$}
\author{Changliang Ren$^{3}$}
\author{Pei Zhao$^{1}$}
\author{Ryosuke Shimizu$^{4}$}
\email{r-simizu@pc.uec.ac.jp}
\author{Pei-Xiang Lu$^{1}$}
\email{lupeixiang@hust.edu.cn}
\affiliation{$^{1}$Laboratory of Optical Information Technology and School of Material Science and Engineering, Wuhan Institute of Technology, Wuhan 430205, China}
\affiliation{$^{2}$Key Laboratory of Low-Dimensional Quantum Structures and Quantum Control of Ministry of Education, Department of Physics and Synergetic Innovation Center for Quantum Effects and Applications, Hunan Normal University, Changsha 410081, China}
\affiliation{$^{3}$Chongqing Institute of Green and Intelligent Technology, Chinese Academy of Sciences, Chongqing 400714, China}
\affiliation{$^{4}$University of Electro-Communications, 1-5-1 Chofugaoka, Chofu, Tokyo 182-8585, Japan}

\date{\today }

\begin{abstract}
Multiparticle interference is a fundamental phenomenon in the study of quantum mechanics.
It was discovered in a recent experiment [Ra, Y.-S. et al, Proc. Natl Acad. Sci. USA \textbf{110}, 1227(2013)] that spectrally uncorrelated biphotons exhibited a nonmonotonic quantum-to-classical transition in a four-photon Hong-Ou-Mandel (HOM) interference. In this work, we consider the same scheme with spectrally correlated photons.
By theoretical calculation and numerical simulation, we found the transition  not only can be nonmonotonic  with negative-correlated or uncorrelated biphotons, but also can be monotonic with  positive-correlated biphotons. The fundamental reason for this difference is that the HOM-type multi-photon interference is a differential-frequency interference.  Our study may shed new light on understanding the role of frequency entanglement in multi-photon behavior.
\end{abstract}



\maketitle

\section{\emph{Introduction}}
Indistinguishability plays an important role in multi-photon interference, which is a fundamental phenomenon in the study of quantum mechanics \cite{Ra2013PNAS, Ou2007, Mosley2008a, Mosley2008b, Jin2013PRA, Jin2011}.
It was believed that, with the increase of indistinguishability, the multi-photon interference pattern changes monotonically \cite{Ra2013PNAS}.
For example, in the case of Hong-Ou-Mandel (HOM) interference demonstrated in 1987 \cite{Hong1987}, the two-fold coincidence counts show a monotonic increase when the time delay
scanned from zero to infinite.
This HOM interference can be interpreted from the viewpoint of indistinguishability: with the increasing of the time delay, the temporal distinguishability (or the decoherence) of the biphoton was also increasing and leading to a quantum-to-classical transition \cite{Ra2013PNAS, Arndt1991, Steck2000}.
Such a monotonic indistinguishability dependence was also observed in  the case of four-photon \cite{Ou1999PRL} and six-photon  \cite{Niu2009} HOM-type interference, where all photons are detected in one output port of the beamsplitter.

\begin{figure*}[tbp]
\centering
\includegraphics[width= 0.85\textwidth]{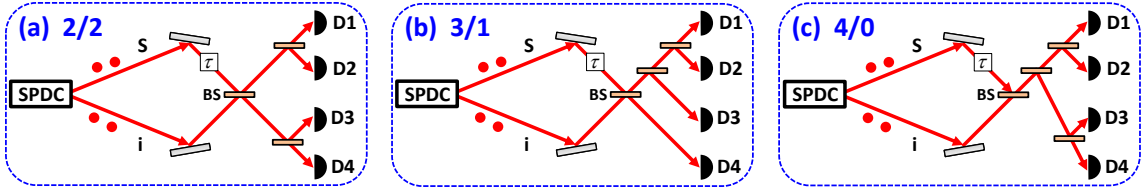}
\caption{(color online)
The 2/2, 3/1 and 4/0 detection schemes for the HOM-type interference. All the beam splitters (BS) are 50:50 beam splitters. $D_n$ ($n$=1, 2, 3, 4) is the single photon detector. $\tau$ is the time delay in the signal arm.}
\label{model}
\end{figure*}

However, recent works  \cite{Ra2013PNAS, Ra2013NC, Tichy2011} evil that such monotonic quantum-to-classical transition was only an exception, i.e., only valid for two-photon cases and for bunching detection in multi-photon cases.
For example, in the four-photon HOM-type experiment \cite{Ra2013PNAS}, where two pair of biphotons were sent to two input ports of a 50:50 beamsplitter and four detectors were prepared at the two output ports (see Fig.\,\ref{model}),
by changing the detection schemes, different interference patterns can be obtained:
in a 2/2 detection (with two detectors at one output port and two detectors at the other port, shown in Fig.\,\ref{model}(a)), the four-fold coincidence counts showed a nonmonotonic indistinguishability dependence;
in contrast, the 4/0 detection scheme %
as shown in Fig.\,\ref{model}(c),
achieved a monotonic dependence.
This study  on the transition between quantum and classical in Ref. \cite{Ra2013PNAS} is important for deeper understanding of the  multi-particle behavior in quantum mechanics.

%
%
%

The interesting phenomenon in Ref.  \cite{Ra2013PNAS}  was realized by spectrally uncorrelated biphotons.  Now a question comes naturally: what phenomenon will be if the biphotons are spectrally correlated? In other words, with the introduction of frequency entanglement, will the interference patterns, especially the monotonicity dependence, be changed?
To answer this question, in this paper, we consider the same  scheme with spectrally correlated (frequency entangled) biphotons.
It will be seen that spectrally correlated biphotons show different interference patterns from the patterns by uncorrelated biphotons. For example, under the 2/2 detection scheme, the spectrally negative- and non-correlated biphotons shows a nonmonotonic dependence, while the  spectrally positively-correlated biphotons shows a monotonic dependence. In contrast, the monotonicity is not affected by the spectral correlation in the 4/0 and 3/1 detection schemes.

This paper is organized as follow:
in the \emph{Introduction} section, we provide the background and motivation of this research.
Then, in the \emph{Theory} section we develop  a multi-mode theory for four-photon HOM-type interference, where the spectral correlation between the signal and idler photons are concerned.
Next, in the \emph{Analysis} section, we first  simulate the HOM-type interference patterns using biphotons with three different spectral correlations: no-correlation, positive-correlation, negative-correlation.
Then, we provide comprehensive discussions on the simulation results.
Finally, we summarize the paper in the \emph{Conclusion} section.
More details for the derivation of the relative equations are given in the \emph{Appendix}.

\section{Theory}
In this paper, we consider a four-photon HOM-type interference with the experimental model shown in Fig.\,\ref{model}.
The four photon state $\left| \psi  \right\rangle $  is generated from the two-pair components in a spontaneous parametric downconversion (SPDC) process.
\begin{equation}\label{eq1}
\begin{array}{lll}
  \left| \psi  \right\rangle & = &  \int_0^\infty d\omega _s d\omega _i d\omega _s^, d\omega _i^,   f(\omega _s ,\omega _i )f(\omega _s^, ,\omega _i^, ) \\
   &  & \times  \hat a_s^\dag  (\omega _s )\hat a_i^\dag  (\omega _i )\hat a_s^\dag  (\omega _s^, )\hat a_i^\dag  (\omega _i^, )\left| {0000} \right\rangle,
\end{array}
\end{equation}
where $\hat{a}^\dagger(\omega)$ is the creation operator at angular frequency
$\omega$, the subscripts $s$ and $i$ denote the signal and idler photons from the first pair,  while $s^,$ and $i^,$ denote the signal and idler photons from the second pair. $f(\omega_s, \omega_i)$ and $f(\omega _s^, ,\omega _i^, )$ are their joint spectral amplitude (JSA).

As calculated in detail in the Appendix, the four-fold coincidence probability $P_{22} (\tau )$ in the 2/2 detection scheme is
\begin{equation}\label{eq2}
\begin{array}{lll}
P_{22} (\tau ) = \frac{1}{{64}}\int_0^\infty  {d\omega _1 d\omega _2 d\omega _3 d\omega _4 } |I_{22}(\tau )|^2,
\end{array}
\end{equation}
with
\begin{equation}\label{eq3}
\begin{array}{lll}
|I_{22}(\tau ) |^2  & = & |(f_{13} f_{24}  + f_{14} f_{23} )e^{ - i\omega _1 \tau } e^{ - i\omega _2 \tau }  \\
             &   &+ (f_{31} f_{42}  + f_{32} f_{41} )e^{ - i\omega _3 \tau } e^{ - i\omega _4 \tau }  \\
             &   &- (f_{12} f_{34}  + f_{14} f_{32} )e^{ - i\omega _1 \tau } e^{ - i\omega _3 \tau }  \\
             &   &- (f_{12} f_{43}  + f_{13} f_{42} )e^{ - i\omega _1 \tau } e^{ - i\omega _4 \tau }  \\
             &   &- (f_{21} f_{34}  + f_{24} f_{31} )e^{ - i\omega _2 \tau } e^{ - i\omega _3 \tau }  \\
             &   &- (f_{21} f_{43}  + f_{23} f_{41} )e^{ - i\omega _2 \tau } e^{ - i\omega _4 \tau }  |^2, \\
\end{array}
\end{equation}
where $f_{mn}=f(\omega _m, \omega _n)$ and $\omega _{m(n)} $ ($m(n)$=1, 2, 3, 4) is the frequency of the detection field for the detectors D$_n$.

The coincidence probability $P_{31} (\tau )$ in the 3/1 detection scheme is
\begin{equation}\label{eq4}
\begin{array}{lll}
P_{31} (\tau ) = \frac{1}{{128}}\int_0^\infty  {d\omega _1 d\omega _2 d\omega _3 d\omega _4 } |I_{31}(\tau )|^2,
\end{array}
\end{equation}
with
\begin{equation}\label{eq5}
\begin{array}{lll}
|I_{31} (\tau )|^2  & = & |-(f_{13} f_{24}  + f_{14} f_{23} )e^{ - i\omega _1 \tau } e^{ - i\omega _2 \tau }  \\
             &   &+ (f_{31} f_{42}  + f_{32} f_{41} )e^{ - i\omega _3 \tau } e^{ - i\omega _4 \tau }  \\
             &   &- (f_{12} f_{34}  + f_{14} f_{32} )e^{ - i\omega _1 \tau } e^{ - i\omega _3 \tau }  \\
             &   &+ (f_{12} f_{43}  + f_{13} f_{42} )e^{ - i\omega _1 \tau } e^{ - i\omega _4 \tau }  \\
             &   &- (f_{21} f_{34}  + f_{24} f_{31} )e^{ - i\omega _2 \tau } e^{ - i\omega _3 \tau }  \\
             &   &+ (f_{21} f_{43}  + f_{23} f_{41} )e^{ - i\omega _2 \tau } e^{ - i\omega _4 \tau }  |^2.
\end{array}
\end{equation}

The coincidence probability $P_{40} (\tau )$ in the 4/0 detection scheme is
\begin{equation}\label{eq6}
\begin{array}{lll}
P_{40} (\tau ) = \frac{1}{{1024}}\int_0^\infty  {d\omega _1 d\omega _2 d\omega _3 d\omega _4 } |I_{40}(\tau )|^2,
\end{array}
\end{equation}
with
\begin{equation}\label{eq7}
\begin{array}{lll}
|I_{40}(\tau ) |^2  & = & |(f_{13} f_{24}  + f_{14} f_{23} )e^{ - i\omega _1 \tau } e^{ - i\omega _2 \tau }  \\
             &   &+ (f_{31} f_{42}  + f_{32} f_{41} )e^{ - i\omega _3 \tau } e^{ - i\omega _4 \tau }  \\
             &   &+ (f_{12} f_{34}  + f_{14} f_{32} )e^{ - i\omega _1 \tau } e^{ - i\omega _3 \tau }  \\
             &   &+ (f_{12} f_{43}  + f_{13} f_{42} )e^{ - i\omega _1 \tau } e^{ - i\omega _4 \tau }  \\
             &   &+ (f_{21} f_{34}  + f_{24} f_{31} )e^{ - i\omega _2 \tau } e^{ - i\omega _3 \tau }  \\
             &   &+ (f_{21} f_{43}  + f_{23} f_{41} )e^{ - i\omega _2 \tau } e^{ - i\omega _4 \tau }  |^2.
\end{array}
\end{equation}

It is interesting to compare the six items in $|I_{22}(\tau )|^2 $,  $|I_{31}(\tau ) |^2 $ and  $|I_{40}(\tau ) |^2 $: the first and second terms in $|I_{22}(\tau ) |^2 $ are positive;  the second, fourth and sixth items in$ |I_{31}(\tau ) |^2$ are positive; all the six items in  $|I_{40}(\tau ) |^2 $ are positive. As calculated in the Appendix, the sign of these terms results from the sign of the transmission and  reflection terms after the beam splitter (BS) in  Fig.\,\ref{model}.
These equations can be further simplified by assuming the exchanging symmetry of $f(\omega _s, \omega _i)=f(\omega _i, \omega _s)$.

\section{Analysis}

\begin{figure}[!tbp]
\centering
\includegraphics[width= 0.48\textwidth]{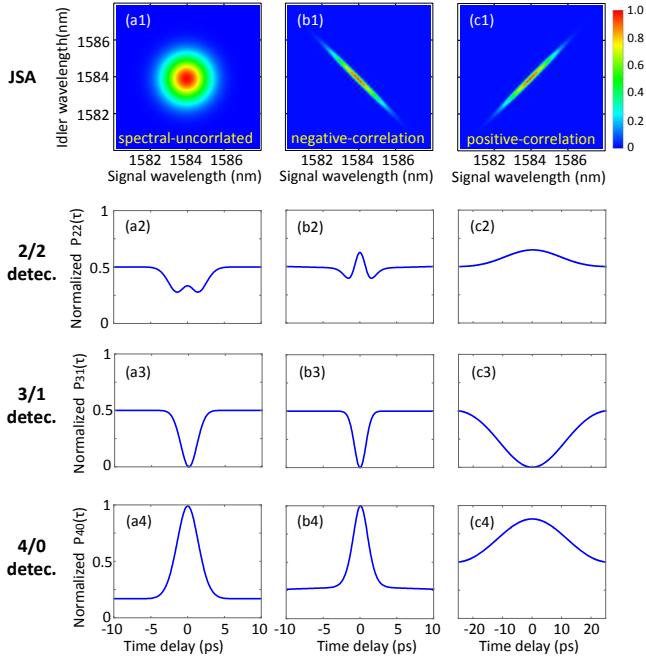}
\caption{(color online)  Three different JSA $f(\omega_s, \omega_i)$: (a1) uncorrelated, (b1) negatively-correlated and (c1) positively-correlated. The corresponding HOM-type interference patterns are shown in (a2-c4):  (a2-c2) are for 2/2 detection scheme; (a3-c3) are for 3/1 detection scheme; (a4-c4) are for 4/0 detection scheme. All the y axes in (a2-c4) are normalized.
}
\label{simulation}
\end{figure}

For a given JSA of $f(\omega_s, \omega_i)$,  using the equations of $P_{22} (\tau )$, $P_{31} (\tau )$ and $P_{40} (\tau )$, it is possible to simulate the HOM-type interference patterns.
Three kinds of JSAs are shown in  Fig.\,\ref{simulation}(a1-c1),  with (a1) spectrally uncorrelated, (b1) positively-correlated and (c1) negatively-correlated. Without the loss of generality, we set the center wavelength of the JSAs at 1584 nm, and set the bandwidth (full width at half maximum) of the signal and idler photons at 2 nm.
Although the shape of the three JSAs is different, the marginal distributions for the signal and idler photons are the same. In other words, from the viewpoint of single photons, all the signal and idler photons have the same spectral distribution in Fig.\,\ref{simulation}(a1-c1).

%

Figure\,\ref{simulation}(a2-c2) show the HOM-type interference patterns for 2/2 detection schemes. It is noteworthy that, for the uncorrelated state (a1) and negatively correlated state (b1), the coincidence probability changes in a nonmonotonic manner, when the time delay changes from 0 to 10 ps.
In contrast, the positively correlated state (c1) shows a monotonic interference pattern.
Figure\,\ref{simulation}(a3-c3) show the HOM-type interference patterns for 3/1 detection schemes, with all the figures in dips, i.e.,  the interference patterns are monotonic when the time delay changes from 0 to infinite.
The patterns for 4/0 detection are shown in Fig.\,\ref{simulation}(a4-c4), with all the figures in bumps, i.e.,  the interference patterns show monotonic dependence.

In Fig.\,\ref{simulation}, biphotons with different correlations show different interference patterns, but what is the underlying physics for such phenomena?
To answer this question, we need to further simplify the Eqs.\,(\ref{eq3}, \ref{eq5}, \ref{eq7}).
As an example, by assuming $f_{mn}=f_{nm}$, Eq.\,(\ref{eq3}) can be simplified as follow.
\begin{equation}\label{eq8}
\begin{array}{l}
|I_{22}(\tau ) |^2  =
(f_{12} f_{34})^2+(f_{13} f_{24})^2+(f_{14} f_{23})^2 \\
+f_{12} f_{13} f_{24} f_{34}+f_{12} f_{14} f_{23} f_{34}+f_{13} f_{14} f_{23} f_{24} \\
+(f_{12} f_{34}+f_{13} f_{24})(f_{12} f_{34}+f_{14} f_{23})\cos(\omega_1-\omega_2) \tau\\
-(f_{12} f_{34}+f_{13} f_{24})(f_{13} f_{24}+f_{14} f_{23})\cos(\omega_1-\omega_3) \tau \\
-(f_{12} f_{34}+f_{14} f_{23})(f_{13} f_{24}+f_{14} f_{23})\cos(\omega_1-\omega_4) \tau\\
-(f_{12} f_{34}+f_{14} f_{23})(f_{13} f_{24}+f_{14} f_{23})\cos(\omega_2-\omega_3) \tau\\
-(f_{12} f_{34}+f_{13} f_{24})(f_{13} f_{24}+f_{14} f_{23})\cos(\omega_2-\omega_4) \tau\\
+(f_{12} f_{34}+f_{13} f_{24})(f_{12} f_{34}+f_{14} f_{23})\cos(\omega_3-\omega_4) \tau\\
+1/2 (f_{13} f_{24}+f_{14} f_{23})^2\cos(\omega_1+\omega_2-\omega_3-\omega_4) \tau\\
+1/2 (f_{12} f_{34}+f_{14} f_{23})^2\cos(\omega_1-\omega_2+\omega_3-\omega_4) \tau\\
+1/2 (f_{12} f_{34}+f_{13} f_{24})^2\cos(\omega_1-\omega_2-\omega_3+\omega_4) \tau.\\
\end{array}
\end{equation}
Obviously, Eq.\,(\ref{eq8}) is a function of $\omega_m-\omega_n$. Similar results can also be derived for Eq.\,(\ref{eq5}) and Eq.\,(\ref{eq7}). So, it can be concluded that \emph{HOM-type multi-photon interference is a differential-frequency interference}. This is true not only for the two-photon HOM interference \cite{Giovannetti2002, Wang2006,   Jin2016QST}, but also for the four-photon HOM interference. Therefore, positively-correlated biphotons, i.e., around $\omega_s-\omega_i=0$,  exhibit different patterns from the one by the uncorrelated biphotons ($\omega_s$ and $\omega_i$ are arbitrary) or negatively-correlated biphotons ($\omega_s+\omega_i=\omega_p$, with $\omega_p$ as the angular frequency of the pump).

The interference patterns in Fig.\,\ref{simulation}(c2-c4) are  ``fatter" (the coherence time is longer) than the patterns in (a2-a4) or (b2-b4). It can also be explained from the above conclusion that HOM type interference is a \emph{differential-frequency} interference.
In fact, Eq.\,(\ref{eq3}) can be viewed as a Fourier transform from frequency-domain to time domain.
Consequently, the  width of the time-domain-interference-pattern is determined by the spectral-domain distribution along the direction of $(\omega _s - \omega _i)$.
The value of $(\omega _s - \omega _i)$ in Fig.\,\ref{simulation}(c1) is the smallest among  (a1-c1) in frequency domain, so the corresponding width in the interference patterns are the largest in time domain, thanks to the spectral positive-correlation in (c1).

It should be emphasized that the theoretical model of our scheme is different from the model in Refs. \cite{Ra2013PNAS, Ra2013NC}, where the spectral correlations are not included.
The photons in the model of Refs. \cite{Ra2013PNAS, Ra2013NC, Ra2017} is spectrally uncorrelated,  therefore, the experiment results in Ref \cite{Ra2013PNAS, Ra2013NC} only correspond to Fig.\,\ref{simulation}(a2, a3, a4) in our simulation.

Many literatures have been dedicated to theoretically analyze the  multi-photon interference using multi-mode theory.
Ou et al analyzed the multi-photon interference using multi-mode theory from spectral modes \cite{Ou1999PRA, Ou1999PRL, Ou2007};
Chen et al modeled the photons as wave packets in time domain \cite{Chen2009};
Ra et al considered Schmidt decomposition on the temporal modes of the photons in their theoretical model  \cite{Ra2013PNAS, Ra2013NC, Tichy2011};
However, in all these theoretical model, the role of spectral correlation is not deeply investigated. To the best of our knowledge, our model is the first theoretical model for multi-photon interference with spectral correlation included.

It is interesting to compare the four-photon HOM interference with the case of the traditional two-photon HOM interference \cite{Hong1987, Jin2015OE}.
The two-fold coincidence probability between two output ports of a beamsplitter (anti-bunching test) can be written as
\begin{equation}\label{eq9}
P_{11} (\tau ) = \frac{1}{4}\int_0^\infty  {d\omega _1 } d\omega _2 |I_{11} (\tau )|^2,
\end{equation}
with
\begin{equation}\label{eq10}
 |I_{11} (\tau )|^2 = | f(\omega _2 ,\omega _1 )e^{ - i\omega _1 \tau }  - f(\omega _1 ,\omega _2 )e^{ - i\omega _2 \tau }|^2
\end{equation}
In contrast, the two-fold coincidence probability of one output port of the beamsplitter (bunching test) can be written as
\begin{equation}\label{eq11}
P_{20} (\tau ) = \frac{1}{16}\int_0^\infty  {d\omega _1 } d\omega _2 |I_{20} (\tau )|^2,
\end{equation}
with
\begin{equation}\label{eq12}
 |I_{20} (\tau )|^2 = | f(\omega _2 ,\omega _1 )e^{ - i\omega _1 \tau }  + f(\omega _1 ,\omega _2 )e^{ - i\omega _2 \tau }|^2
\end{equation}
We also simulated  $P_{11}(\tau )$  and $P_{20}(\tau )$  using the three JSAs shown in Fig.\,\ref{simulation}(a1-c1). It was found that the monotonicity was not affect by the spectral correlations, i.e., all the three  $P_{11}(\tau )$ patterns  show dips, while all the three  $P_{20}(\tau )$ patterns  show bumps for the JSAs in Fig.\,\ref{simulation}(a1-c1).

It is also important to rethink the prerequisite condition for 100 \% visibility in the two-photon and four-photon HOM interference.
In the two-photon case, \emph{exchanging symmetry} of $f(\omega _1 ,\omega _2 )=f(\omega _2 ,\omega _1 )$ is required to achieve 100 \% visibility, i.e.,  $P_{11} (0 )=0$ \cite{Shimizu2009, Jin2015OE}. In contrast, the prerequisite condition is complex for the four-photon HOM interference to achieve 100 \% visibility. For example, in the case of 3/1 detection, $P_{31} (0)=0$ implies
$-(f_{13} f_{24}  + f_{14} f_{23} )
+ (f_{31} f_{42}  + f_{32} f_{41} )
- (f_{12} f_{34}  + f_{14} f_{32} )
+ (f_{12} f_{43}  + f_{13} f_{42} )
- (f_{21} f_{34}  + f_{24} f_{31} )
+ (f_{21} f_{43}  + f_{23} f_{41} )=0$, which is an upgraded version of the \emph{exchanging symmetry} for the four-photon case.

In the theoretical model in Eq.\,(\ref{eq1}), the four-photon state is generated from a double pair emission, which has a spectral distribution of $f(\omega _s ,\omega _i )f(\omega _s^, ,\omega _i^, )$. In the future, it is possible to directly generate a four-photon state with a spectral distribution of $f(\omega _s ,\omega _i, \omega _s^, ,\omega _i^, )$.  This state may be generated from, say, a fourth-order spontaneous parametric down conversion process, where a higher-energy photon ``splits'' into four lower-energy photons. For example, a 1600 nm photon may be downconverted to four 400 nm photons.  This is the inverse process of a fourth harmonic generation. The direct generation of three-photon state has been chased by several groups for a long time \cite{Bouwmeester1999, Ren2012, Shalm2013, Agne2016}. It is also interesting to study the case of four-photon state  \cite{Riedmatten2004, Wang2015, Hiesmayr2016}. In this case, the spectral correlations and the HOM interference might be different from the case discussed in this paper. It will be a interesting topic to investigate in the future.
Another future work is to expand the theoretical model to the case of six-photon and more photons. Although the equations might be complex, the expansion method is direct, i.e., similar as what we did in the work.

For the future experimental demonstration, our scheme has been ready to be realized with the state-of-art technologies. The spectrally uncorrelated JSA in Fig.\,\ref{simulation}(a1) can be generated by  filtering a PPKTP downconversion source at 1584 nm \cite{Jin2013OE, Bruno2014, Jin2016PRAppl}.
The spectrally negatively correlated JSA in Fig.\,\ref{simulation}(b1) has been generated in a ps-pulse-pumped PPSLT crystal \cite{Shimizu2009}, while the  spectrally positively correlated JSA in Fig.\,\ref{simulation}(c1) has been prepared in a fs-pulse-pumped PPKTP crystal \cite{Shimizu2009}. For detection, we can use the similar setup demonstrated recently \cite{Jin2016SR}.

Our work have several applications in the future. Higher-order correlations in many-body system are very important for characterizing a quantum system and became to be a hot topic in study of quantum optics \cite{Schweigler2017, Ra2017, Tichy2014}.  In this work, we studied the role of spectral correlation  in a  four-photon quantum interference,  which actually corresponds to a  fourth-order temporal correlation in a four-body system. Therefore, this work may make contribution to the deep understanding of higher-order correlations of a quantum system. Another possible application of our work is for quantum sensing based on Hong-Ou-Mandel interference \cite{Esfahani2015, Hou2014, Liu2016} . Thirdly, the spectral correlation may be applied to the reduction of detection noise in a dispersive medium, which has been recently demonstrate in Ref \cite{Sedziak2017} with only two photons.   In the case of four photons, the noise-reduction effect might be enhanced.

\section{Conclusion}
In conclusion, we have investigated the role of spectral correlation (frequency entanglement) in quantum-to-classical transition in a four-photon Hong-Ou-Mandel interference.
By theoretical calculation and numerical simulation based on a multi-mode theory for spectrally correlated photons, it was found that the transition can be monotonic for positively-correlated biphotons, and can be nonmonotonic negative-, or non-correlated biphotons in the 2/2 detection scheme. In contrast, the monotonicity  was not changed in the 3/1 and 4/0 detection schemes. The fundamental reason for these difference is: the HOM-type interference is a differential-frequency interference. Our theoretical scheme can be easily demonstrated in experiment using the state-of-art technologies. This study may shed new light on understanding the role of entanglement in multi-photon behavior.

\section*{Acknowledgements}
The authors are grateful to M. Takeoka for helpful discussions.  
R.-B. J. is supported by Fund from the Educational Department of Hubei Province, China (Grant No. D20161504).
C. L. R. is supported by Youth Innovation Promotion Association (CAS) No.2015317, National Natural Science Foundations of China (Grant No.11605205), Natural Science Foundations of Chong Qing (No.cstc2015jcyjA00021).
H. J. is supported by National Natural Science Foundations of China (Grant No. 11474087).


\clearpage

\section*{Appendix}
Here we deduce the equations for the four photon Hong-Ou-Mandel (HOM) type interference in detail.
The setup of HOM interference with 2/2 detection scheme is shown in  Fig.\,\ref{model}(a).
The two-pair component  from a spontaneous parametric down conversion (SPDC) process is expressed as  (Eq.\,(\ref{eq1}) in the main text).
\begin{equation}\label{eqA1}
\begin{array}{lll}
  \left| \psi  \right\rangle & = &  \int_0^\infty d\omega _s d\omega _i d\omega _s^, d\omega _i^,   f(\omega _s ,\omega _i )f(\omega _s^, ,\omega _i^, ) \\
   &  & \times  \hat a_s^\dag  (\omega _s )\hat a_i^\dag  (\omega _i )\hat a_s^\dag  (\omega _s^, )\hat a_i^\dag  (\omega _i^, )\left| {0000} \right\rangle,
\end{array}
\end{equation}
The meaning of each parameter are explained in the main text.
The detection field  operator of detector $D_n$ ($n$=1, 2, 3, 4) is
\begin{equation}\label{eqA2}
\hat E_n^{( + )} (t_n ) = \frac{1}{{\sqrt {2\pi } }}\int_0^\infty  {d\omega _n \hat a_n (\omega _n )} e^{ - i\omega _n t_n },
\end{equation}
where $\omega_n$   is the frequency of the detection field.  $\hat a_n$ is the annihilation operator of the detection field.
The transformation rule of a 50/50 beamsplitter  is
$
\hat a_{o_1}  = \frac{1}{{\sqrt 2 }}(\hat a_{in_1}  + \hat a_{in_2} )$  and $\hat a_{o_2}  = \frac{1}{{\sqrt 2 }}(\hat a_{in_1}  - \hat a_{in_2} )$,
where the subscripts $o_1$ and $o_2$ denote the two output ports of the beamsplitter, while the $in_1$ and $in_2$ denote the two input ports.

So, we can write the detection fields as
\begin{equation}\label{eqA3}
\begin{array}{l}
 \hat E_1^{( + )} (t_1 ) = \frac{1}{{2\sqrt {2\pi } }}\int_0^\infty  {d\omega _1 } [\hat a_s (\omega _1 )e^{ - i\omega _1 \tau }  + \hat a_i (\omega _1 )]e^{ - i\omega _1 t_1 } , \\
 \hat E_2^{( + )} (t_2 ) = \frac{1}{{2\sqrt {2\pi } }}\int_0^\infty  {d\omega _2 } [\hat a_s (\omega _2 )e^{ - i\omega _2 \tau }  + \hat a_i (\omega _2 )]e^{ - i\omega _2 t_2 } , \\
 \hat E_3^{( + )} (t_3 ) = \frac{1}{{2\sqrt {2\pi } }}\int_0^\infty  {d\omega _3 } [\hat a_s (\omega _3 )e^{ - i\omega _3 \tau }  - \hat a_i (\omega _3 )]e^{ - i\omega _3 t_3 } , \\
 \hat E_4^{( + )} (t_4 ) = \frac{1}{{2\sqrt {2\pi } }}\int_0^\infty  {d\omega _4 } [\hat a_s (\omega _4 )e^{ - i\omega _4 \tau }  - \hat a_i (\omega _4 )]e^{ - i\omega _4 t_4 } ,
 \end{array}
\end{equation}
where the phase term $e^{ - i\omega _n \tau }$  is introduced by the time delay  $\tau$.
The coincidence probability $P_{22}$  as a function of delay time $\tau$   can be expressed as
\begin{equation}\label{eqA4}
\begin{array}{l}
P_{22} (\tau ) =  \int {dt_1 dt_2 } dt_3 dt_4 \times \\
 \left\langle {\psi \left| {\hat E_4^{( - )} \hat E_3^{( - )} \hat E_2^{( - )} \hat E_1^{( - )} \hat E_1^{( + )} \hat E_2^{( + )} \hat E_3^{( + )} \hat E_4^{( + )} } \right|\psi } \right\rangle.
 \end{array}
\end{equation}
First, let us consider the $\hat E_1^{( + )} \hat E_2^{( + )} \hat E_3^{( + )} \hat E_4^{( + )} \left| \psi  \right\rangle$. For simplicity, the key components can be  written as: $[\hat a_s (\omega _1 ) + \hat a_i (\omega _1 )][\hat a_s (\omega _2 ) + \hat a_i (\omega _2 )][\hat a_s (\omega _3 ) - \hat a_i (\omega _3 )][\hat a_s (\omega _4 ) - \hat a_i (\omega _4 )]$.
Only 6 out of 16 terms exist:
$\hat a_s \hat a_s \hat a_i \hat a_i $, $\hat a_i \hat a_i \hat a_s \hat a_s $,  $- \hat a_s \hat a_i \hat a_s \hat a_i $,  $- \hat a_s \hat a_i \hat a_i \hat a_s $, $- \hat a_i \hat a_s \hat a_s \hat a_i $ and $- \hat a_i \hat a_s \hat a_i \hat a_s$.
The first term ($\hat a_s \hat a_s \hat a_i \hat a_i$) is
\begin{widetext}
\begin{spacing}{1.6}
\begin{equation}\label{eqA5}
\begin{array}{l}
 \frac{1}{{16}}(\frac{1}{{2\pi }})^2 \int_0^\infty  {d\omega _1 d\omega _2 d\omega _3 d\omega _4 } \hat a_s (\omega _1 )\hat a_s (\omega _2 )\hat a_i (\omega _3 )\hat a_i (\omega _4 )e^{ - i\omega _1 \tau } e^{ - i\omega _2 \tau } e^{ - i\omega _1 t_1 } e^{ - i\omega _2 t_2 } e^{ - i\omega _3 t_3 } e^{ - i\omega _4 t_4 }  \\
  \times \int_0^\infty  {d\omega _s d\omega _i d\omega _{s^, } d\omega _{i^, } } f(\omega _s ,\omega _i )f(\omega _s^, ,\omega _i^, )\hat a_s^\dag  (\omega _s )\hat a_i^\dag  (\omega _i )\hat a_s^\dag  (\omega _s^, )\hat a_i^\dag  (\omega _i^, )\left| 0 \right\rangle  \\
 {\rm{ = }}\frac{1}{{16}}(\frac{1}{{2\pi }})^2 \int_0^\infty  {d\omega _1 d\omega _2 d\omega _3 d\omega _4 } \int_0^\infty  {d\omega _s d\omega _i d\omega _{s^, } d\omega _{i^, } } [\delta (\omega _1  - \omega _s )\delta (\omega _2  - \omega _{s^, } ) + \delta (\omega _1  - \omega _{s^, } )\delta (\omega _2  - \omega _s )] [\delta (\omega _3  - \omega _i )\\
  \delta (\omega _4  - \omega _{i^, } ) + \delta (\omega _3  - \omega _{i^, } )\delta (\omega _4  - \omega _i )]f(\omega _s ,\omega _i )f(\omega _s^, ,\omega _i^, )e^{ - i\omega _1 \tau } e^{ - i\omega _2 \tau } e^{ - i\omega _1 t_1 } e^{ - i\omega _2 t_2 } e^{ - i\omega _3 t_3 } e^{ - i\omega _4 t_4 } \left| 0 \right\rangle  \\
 {\rm{ = }}\frac{1}{{16}}(\frac{1}{{2\pi }})^2 \int_0^\infty  {d\omega _1 d\omega _2 d\omega _3 d\omega _4 } [f(\omega _1 ,\omega _3 )f(\omega _2 ,\omega _4 ) + f(\omega _1 ,\omega _4 )f(\omega _2 ,\omega _3 ) + f(\omega _2 ,\omega _3 )f(\omega _1 ,\omega _4 ) +  \\
 f(\omega _2 ,\omega _4 )f(\omega _1 ,\omega _3 )]e^{ - i\omega _1 \tau } e^{ - i\omega _2 \tau } e^{ - i\omega _1 t_1 } e^{ - i\omega _2 t_2 } e^{ - i\omega _3 t_3 } e^{ - i\omega _4 t_4 } \left| 0 \right\rangle  \\
 {\rm{ = }}\frac{1}{8}(\frac{1}{{2\pi }})^2 \int_0^\infty  {d\omega _1 d\omega _2 d\omega _3 d\omega _4 } [f(\omega _1 ,\omega _3 )f(\omega _2 ,\omega _4 ) + f(\omega _1 ,\omega _4 )f(\omega _2 ,\omega _3 )]e^{ - i\omega _1 \tau } e^{ - i\omega _2 \tau } e^{ - i\omega _1 t_1 } e^{ - i\omega _2 t_2 } e^{ - i\omega _3 t_3 } e^{ - i\omega _4 t_4 } \left| 0 \right\rangle  \\
 {\rm{ = }}\frac{1}{8}(\frac{1}{{2\pi }})^2 \int_0^\infty  {d\omega _1 d\omega _2 d\omega _3 d\omega _4 } \times ff_1 \times e^{ - i\omega _1 t_1 } e^{ - i\omega _2 t_2 } e^{ - i\omega _3 t_3 } e^{ - i\omega _4 t_4 } \left| 0 \right\rangle
 \end{array}
\end{equation}
\end{spacing}
\end{widetext}
where
 \begin{spacing}{1.6}
 \begin{equation}\label{eqA6}
\begin{array}{lll}
ff_1 &=&[f(\omega _1 ,\omega _3 )f(\omega _2 ,\omega _4 ) + f(\omega _1 ,\omega _4 )f(\omega _2 ,\omega _3 )] \\
     & &   e^{ - i\omega _1 \tau } e^{ - i\omega _2 \tau }.
\end{array}
\end{equation}
\end{spacing}
In the above calculation, the following relationship is used.
\begin{spacing}{1.6}
\begin{equation}\label{eqA7}
\begin{array}{l}
 \hat a_s (\omega _1 )\hat a_s (\omega _2 )\hat a_s^\dag  (\omega _s )\hat a_s^\dag  (\omega _s^, )\left| 0 \right\rangle  \\ =
 [ \delta (\omega _1  - \omega _s )\delta (\omega _2  - \omega _{s^, } ) + \delta (\omega _1  - \omega _{s^, } )\delta (\omega _2  - \omega _s )] \left| 0 \right\rangle
 \end{array}
\end{equation}
\end{spacing}
Similarly, the second term ($\hat a_i \hat a_i \hat a_s \hat a_s$) is£º
\begin{spacing}{1.6}
\begin{equation}\label{eqA8}
\begin{array}{l}
\frac{1}{8}(\frac{1}{{2\pi }})^2 \int_0^\infty  {d\omega _1 d\omega _2 d\omega _3 d\omega _4 }  \times ff_2  \times  e^{ - i\omega _1 t_1 } e^{ - i\omega _2 t_2 }  \\ e^{ - i\omega _3 t_3 } e^{ - i\omega _4 t_4 } \left| 0 \right\rangle,
\end{array}
\end{equation}
\end{spacing}
where
\begin{spacing}{1.6}
\begin{equation} \label{eqA9}
\begin{array}{lll}
ff_2 &=&  [ f(\omega _3 ,\omega _1 )f(\omega _4 ,\omega _2 ,) + f(\omega _3 ,\omega _2 )f(\omega _4 ,\omega _1 ) ]\\
     & &   e^{ - i\omega _3 \tau }  e^{ - i\omega _4 \tau }.
\end{array}
\end{equation}
\end{spacing}
The third term ($- \hat a_s \hat a_i \hat a_s \hat a_i$) is
\begin{spacing}{1.6}
\begin{equation}\label{eqA10}
\begin{array}{l}
\frac{1}{8}(\frac{1}{{2\pi }})^2 \int_0^\infty  {d\omega _1 d\omega _2 d\omega _3 d\omega _4 }  \times ff_3  \times e^{ - i\omega _1 t_1 }  e^{ - i\omega _2 t_2 } \\ e^{ - i\omega _3 t_3 } e^{ - i\omega _4 t_4 } \left| 0 \right\rangle, \\
\end{array}
\end{equation}
\end{spacing}
where
\begin{spacing}{1.6}
\begin{equation} \label{eqA11}
\begin{array}{lll}
ff_3 &=& [ - f(\omega _1 ,\omega _2 )f(\omega _3 ,\omega _4 ) - f(\omega _1 ,\omega _4 )f(\omega _3 ,\omega _2 )] \\
     & &  e^{ - i\omega _1 \tau } e^{ - i\omega _3 \tau }. \\
\end{array}
\end{equation}
\end{spacing}
The fourth term ($- \hat a_s \hat a_i \hat a_i \hat a_s$) is
\begin{spacing}{1.6}
\begin{equation}\label{eqA12}
\begin{array}{l}
 \frac{1}{8}(\frac{1}{{2\pi }})^2 \int_0^\infty  {d\omega _1 d\omega _2 d\omega _3 d\omega _4  \times } ff_4  \times e^{ - i\omega _1 t_1 } e^{ - i\omega _2 t_2 }\\ e^{ - i\omega _3 t_3 } e^{ - i\omega _4 t_4 } \left| 0 \right\rangle,\\
\end{array}
\end{equation}
\end{spacing}
where
\begin{spacing}{1.6}
\begin{equation}\label{eqA13}
\begin{array}{lll}
 ff_4  &=& [ - f(\omega _1 ,\omega _2 )f(\omega _4 ,\omega _3 ) - f(\omega _1 ,\omega _3 )f(\omega _4 ,\omega _2 )]\\
       & & e^{ - i\omega _1 \tau } e^{ - i\omega _4 \tau }.
\end{array}
\end{equation}
\end{spacing}
The fifth term ($- \hat a_i \hat a_s \hat a_s \hat a_i$) is:
\begin{spacing}{1.6}
\begin{equation}\label{eqA14}
\begin{array}{l}
\frac{1}{8}(\frac{1}{{2\pi }})^2 \int_0^\infty  {d\omega _1 d\omega _2 d\omega _3 d\omega _4 }  \times ff_5  \times e^{ - i\omega _1 t_1 } e^{ - i\omega _2 t_2 } \\  e^{ - i\omega _3 t_3 } e^{ - i\omega _4 t_4 } \left| 0 \right\rangle,
\end{array}
\end{equation}
\end{spacing}
where
\begin{spacing}{1.6}
\begin{equation}\label{eqA15}
\begin{array}{lll}
ff_5  &=& [ - f(\omega _2 ,\omega _1 )f(\omega _3 ,\omega _4 ) - f(\omega _2 ,\omega _4 )f(\omega _3 ,\omega _1 )]\\
& & e^{ - i\omega _2 \tau } e^{ - i\omega _3 \tau }.
\end{array}
\end{equation}
\end{spacing}
The sixth term ($- \hat a_i \hat a_s \hat a_i \hat a_s$) is:
\begin{spacing}{1.6}
\begin{equation}\label{eqA16}
\begin{array}{l}
\frac{1}{8}(\frac{1}{{2\pi }})^2 \int_0^\infty  {d\omega _1 d\omega _2 d\omega _3 d\omega _4 }  \times ff_6  \times e^{ - i\omega _1 t_1 } e^{ - i\omega _2 t_2 } \\ e^{ - i\omega _3 t_3 } e^{ - i\omega _4 t_4 } \left| 0 \right\rangle,
\end{array}
\end{equation}
\end{spacing}
where
\begin{spacing}{1.6}
\begin{equation}\label{eqA17}
\begin{array}{lll}
ff_6  &=& [ - f(\omega _2 ,\omega _1 )f(\omega _4 ,\omega _3 ) - f(\omega _2 ,\omega _3 )f(\omega _4 ,\omega _1 )]\\
     & & e^{ - i\omega _2 \tau } e^{ - i\omega _4 \tau }.
\end{array}
\end{equation}
\end{spacing}
Combine these six terms:
\begin{spacing}{1.6}
\begin{equation}\label{eqA18}
\begin{array}{l}
 \hat E_1^{( + )} \hat E_2^{( + )} \hat E_3^{( + )} \hat E_4^{( + )} \left| \psi  \right\rangle  =  \\  \frac{1}{8}(\frac{1}{{2\pi }})^2 \int_0^\infty  {d\omega _1 d\omega _2 d\omega _3 d\omega _4 } (ff_1
  + ff_2  + ff_3  + ff_4  \\  + ff_5  + ff_6 )     e^{ - i\omega _1 t_1 } e^{ - i\omega _2 t_2 } e^{ - i\omega _3 t_3 } e^{ - i\omega _4 t_4 } \left| 0 \right\rangle.
 \end{array}
\end{equation}
\end{spacing}
Then
\begin{spacing}{1.6}
\begin{equation}\label{eqA19}
\begin{array}{l}
 \left\langle {\psi \left| {\hat E_4^{( - )} \hat E_3^{( - )} \hat E_2^{( - )} \hat E_1^{( - )} \hat E_1^{( + )} \hat E_2^{( + )} \hat E_3^{( + )} \hat E_4^{( + )} } \right|\psi } \right\rangle  \\
  = \frac{1}{8}(\frac{1}{{2\pi }})^2 \int_0^\infty  {d\omega _1 d\omega _2 d\omega _3 d\omega _4 } (ff_1  + ff_2  + ff_3  + ff_4  \\
  + ff_5  + ff_6 )e^{ - i\omega _1 t_1 } e^{ - i\omega _2 t_2 } e^{ - i\omega _3 t_3 } e^{ - i\omega _4 t_4 }  \times  \\
 \frac{1}{8}(\frac{1}{{2\pi }})^2 \int_0^\infty  {d\omega _1^, d\omega _2^, d\omega _3^, d\omega _4^, } (ff_1^*  + ff_2^*  + ff_3^*  + ff_4^*  \\
  + ff_5^*  + ff_6^* )e^{ i\omega _1^, t_1 } e^{ i\omega _2^, t_2 } e^{ i\omega _3^, t_3 } e^{ i\omega _4^, t_4 },
 \end{array}
\end{equation}
\end{spacing}
where, $ff^*$ is the complex conjugate of $ff$.

Finally,
\begin{widetext}
\begin{spacing}{1.6}
\begin{equation}\label{eqA20}
\begin{array}{lll}
 P_{22} (\tau ) &=& \int {dt_1 dt_2 } dt_3 dt_4 \left\langle {\psi \left| {\hat E_4^{( - )} \hat E_3^{( - )} \hat E_2^{( - )} \hat E_1^{( - )} \hat E_1^{( + )} \hat E_2^{( + )} \hat E_3^{( + )} \hat E_4^{( + )} } \right|\psi } \right\rangle  \\
  &=& \frac{1}{{64}}(\frac{1}{{2\pi }})^4 \int {dt_1 dt_2 } dt_3 dt_4 \int_0^\infty  {d\omega _1 d\omega _2 d\omega _3 d\omega _4 } \int_0^\infty  {d\omega _1^, d\omega _2^, d\omega _3^, d\omega _4^, } (ff_1  + ff_2  + ff_3  + ff_4  + ff_5  + ff_6 ) \\
  & &(ff_1^*  + ff_2^*  + ff_3^*  + ff_4^*  + ff_5^*  + ff_6^* )e^{ - i\omega _1 t_1 } e^{ - i\omega _2 t_2 } e^{ - i\omega _3 t_3 } e^{ - i\omega _4 t_4 } e^{i\omega _1^, t_1 } e^{i\omega _2^, t_2 } e^{i\omega _3^, t_3 } e^{i\omega _4^, t_4 }  \\
  &=& \frac{1}{{64}}(\frac{1}{{2\pi }})^4 \int_0^\infty  {d\omega _1 d\omega _2 d\omega _3 d\omega _4 } \int_0^\infty  {d\omega _1^, d\omega _2^, d\omega _3^, d\omega _4^, } (ff_1  + ff_2  + ff_3  + ff_4  + ff_5  + ff_6 ) \\
  & &(ff_1^*  + ff_2^*  + ff_3^*  + ff_4^*  + ff_5^*  + ff_6^* )(2\pi )^4 \delta (\omega _1  - \omega _1^, )\delta (\omega _2  - \omega _2^, )\delta (\omega _3  - \omega _3^, )\delta (\omega _4  - \omega _4^, ) \\
  &=& \frac{1}{{64}}\int_0^\infty  {d\omega _1 d\omega _2 d\omega _3 d\omega _4 } \left| {ff_1  + ff_2  + ff_3  + ff_4  + ff_5  + ff_6 } \right|^2
 \end{array}
\end{equation}
\end{spacing}
\end{widetext}
In the above calculation, the relationship of $\delta (\omega  - \omega ^, ) = \frac{1}{{2\pi }}\int_{ - \infty }^\infty  {e^{i(\omega  - \omega ^, )t} } dt$ is used;

In conclusion, the four-fold coincidence probability in the 2/2 detection scheme is
\begin{equation}\label{eqA21}
P_{22} (\tau ) = \frac{1}{{64}}\int_0^\infty  {d\omega _1 d\omega _2 d\omega _3 d\omega _4 } | I_{22}(\tau ) |^{\rm{2}},
\end{equation}
with
\begin{equation}\label{eqA22}
\begin{array}{lll}
|I_{22} (\tau )|^2  & = & |(f_{13} f_{24}  + f_{14} f_{23} )e^{ - i\omega _1 \tau } e^{ - i\omega _2 \tau }  \\
             &   &+ (f_{31} f_{42}  + f_{32} f_{41} )e^{ - i\omega _3 \tau } e^{ - i\omega _4 \tau }  \\
             &   &- (f_{12} f_{34}  + f_{14} f_{32} )e^{ - i\omega _1 \tau } e^{ - i\omega _3 \tau }  \\
             &   &- (f_{12} f_{43}  + f_{13} f_{42} )e^{ - i\omega _1 \tau } e^{ - i\omega _4 \tau }  \\
             &   &- (f_{21} f_{34}  + f_{24} f_{31} )e^{ - i\omega _2 \tau } e^{ - i\omega _3 \tau }  \\
             &   &- (f_{21} f_{43}  + f_{23} f_{41} )e^{ - i\omega _2 \tau } e^{ - i\omega _4 \tau }  |^2,
\end{array}
\end{equation}
where $f_{ij}=f(\omega _i, \omega _j)$.

In the 3/1 detection, the key components can be written as
$
[\hat a_s (\omega _1 ) + \hat a_i (\omega _1 )][\hat a_s (\omega _2 ) + \hat a_i (\omega _2 )][\hat a_s (\omega _3 ) + \hat a_i (\omega _3 )][\hat a_s (\omega _4 ) - \hat a_i (\omega _4 )]
$.
Only 6 out of 16 terms exist:
 $- \hat a_s \hat a_s \hat a_i \hat a_i$,
$\hat a_i \hat a_i \hat a_s \hat a_s$,
$ - \hat a_s \hat a_i \hat a_s \hat a_i$,
$\hat a_s \hat a_i \hat a_i \hat a_s$,
$ - \hat a_i \hat a_s \hat a_s \hat a_i$ and
$\hat a_i \hat a_s \hat a_i \hat a_s $.
Following the similar method as in the case of 2/2 detection,
the coincidence probability $P_{31} (\tau )$ in the 3/1 detection scheme can be calculated as
\begin{equation}\label{eqA23}
\begin{array}{lll}
P_{31} (\tau ) = \frac{1}{{128}}\int_0^\infty  {d\omega _1 d\omega _2 d\omega _3 d\omega _4 } |I_{31}(\tau )|^2,
\end{array}
\end{equation}
with
\begin{equation}\label{eqA24}
\begin{array}{lll}
|I_{31}(\tau ) |^2  & = & |-(f_{13} f_{24}  + f_{14} f_{23} )e^{ - i\omega _1 \tau } e^{ - i\omega _2 \tau }  \\
             &   &+ (f_{31} f_{42}  + f_{32} f_{41} )e^{ - i\omega _3 \tau } e^{ - i\omega _4 \tau }  \\
             &   &- (f_{12} f_{34}  + f_{14} f_{32} )e^{ - i\omega _1 \tau } e^{ - i\omega _3 \tau }  \\
             &   &+ (f_{12} f_{43}  + f_{13} f_{42} )e^{ - i\omega _1 \tau } e^{ - i\omega _4 \tau }  \\
             &   &- (f_{21} f_{34}  + f_{24} f_{31} )e^{ - i\omega _2 \tau } e^{ - i\omega _3 \tau }  \\
             &   &+ (f_{21} f_{43}  + f_{23} f_{41} )e^{ - i\omega _2 \tau } e^{ - i\omega _4 \tau }  |^2.
\end{array}
\end{equation}

In the 4/0 detection, the key components can be written as
$
[\hat a_s (\omega _1 ) + \hat a_i (\omega _1 )][\hat a_s (\omega _2 ) + \hat a_i (\omega _2 )][\hat a_s (\omega _3 ) + \hat a_i (\omega _3 )][\hat a_s (\omega _4 ) + \hat a_i (\omega _4 )]
$.
Only 6 out of 16 terms exist:
 $\hat a_s \hat a_s \hat a_i \hat a_i$,
$\hat a_i \hat a_i \hat a_s \hat a_s$,
$\hat a_s \hat a_i \hat a_s \hat a_i$,
$\hat a_s \hat a_i \hat a_i \hat a_s$,
$\hat a_i \hat a_s \hat a_s \hat a_i$ and
$\hat a_i \hat a_s \hat a_i \hat a_s $.
The coincidence probability $P_{40} (\tau )$ in the 4/0 detection scheme is
\begin{equation}\label{eqA25}
\begin{array}{lll}
P_{40} (\tau ) = \frac{1}{{1024}}\int_0^\infty  {d\omega _1 d\omega _2 d\omega _3 d\omega _4 } |I_{40}(\tau )|^2,
\end{array}
\end{equation}
with
\begin{equation}\label{eqA26}
\begin{array}{lll}
|I_{40}(\tau ) |^2  & = & |(f_{13} f_{24}  + f_{14} f_{23} )e^{ - i\omega _1 \tau } e^{ - i\omega _2 \tau }  \\
             &   &+ (f_{31} f_{42}  + f_{32} f_{41} )e^{ - i\omega _3 \tau } e^{ - i\omega _4 \tau }  \\
             &   &+ (f_{12} f_{34}  + f_{14} f_{32} )e^{ - i\omega _1 \tau } e^{ - i\omega _3 \tau }  \\
             &   &+ (f_{12} f_{43}  + f_{13} f_{42} )e^{ - i\omega _1 \tau } e^{ - i\omega _4 \tau }  \\
             &   &+ (f_{21} f_{34}  + f_{24} f_{31} )e^{ - i\omega _2 \tau } e^{ - i\omega _3 \tau }  \\
             &   &+ (f_{21} f_{43}  + f_{23} f_{41} )e^{ - i\omega _2 \tau } e^{ - i\omega _4 \tau }  |^2.
\end{array}
\end{equation}

\end{document}